\documentclass[doublespacing]{elsart}

\usepackage{epsfig}

\usepackage{amssymb}

\begin{document}

\begin{frontmatter}



\title{Stochastic Model of Yeast Cell Cycle Network}


\author[1,2]{Yuping Zhang}
\author[1,2]{Minping Qian\corauthref{cor1}}
\ead{qianmp@math.pku.edu.cn} \corauth[cor1]{Corresponding author
at: Center for Theoretical Biology and School of Mathematical
Sciences Peking University- Beijing, 100871, China. Phone: (8610)
6275-2525 Fax: (8610) 6275-9595}
\author[1,3]{Qi Ouyang\corauthref{cor2}}
\ead{qi@pku.edu.cn} \corauth[cor2]{Corresponding author at: Center
for Theoretical Biology and Department of Physics Peking
University- Beijing, 100871, China. Phone: (8610) 6275-6943 Fax:
(8610) 6275-9595}
\author[1,2]{Minghua Deng}
\author[1,3]{Fangting Li}
\author[1,4]{Chao Tang}

\address[1]{Center for Theoretical Biology Peking University- Beijing, 100871, China}
\address[2]{School of Mathematical Sciences Peking University- Beijing, 100871, China}
\address[3]{Department of Physics Peking University- Beijing, 100871, China}
\address[4]{California Institute for Quantitative Biomedical Research, \\
 Departments of Biopharmaceutical Sciences and Biochemistry and Biophysics
 \\ - UCSF Box 2540, University of California at San Francisco, San Francisco, CA 94143-2540}

\begin{abstract}
Biological functions in living cells are controlled by protein
interaction and genetic networks. These molecular networks should be
dynamically stable against various fluctuations which are inevitable
in the living world. In this paper, we propose and study a
stochastic model for the network regulating the cell cycle of the
budding yeast. The stochasticity in the model is controlled by a
temperature-like parameter $\beta$. Our simulation results show that
both the biological stationary state and the biological pathway are
stable for a wide range of ``temperature". There is, however, a
sharp transition-like behavior at $\beta_c$, below which the
dynamics is dominated by noise. We also define a pseudo energy
landscape for the system in which the biological pathway can be seen
as a deep valley.
\end{abstract}

\begin{keyword}
protein network \sep Markov chain \sep stochastic \sep dynamic

\PACS 87.17.Aa \sep 87.17.-d \sep 87.10.+e \sep 87.16.Yc
\end{keyword}
\end{frontmatter}

\section{Introduction}
\label{Introduction} Quantitative understanding of biological
systems and functions from their components and interactions
presents a challenge as well as an opportunity for interested
scientists of various disciplines. Recently, a considerable amount
of attention has been paid to the quantitative modeling and
understanding of the budding yeast cell cycle regulation
\cite{Chen2000,Cross2002,Li2004,ChenHC2004,ChenKC2004,Cross2005,Futcher2002,Murray2004,Ingolia2004,Tyers2004}.
In particular, Li \textit{et al.} \cite{Li2004} introduced a
deterministic Boolean network model and investigated its dynamic
and structural properties. Their main results are that the network
is both dynamically and structurally stable. The biological
stationary state is a global attractor of the dynamics; the
biological pathway is a globally attracting dynamic trajectory.
These properties are largely preserved with respect to small
structural perturbations to the network, e.g. adding or deleting
links. However, one crucial point left unaddressed in their study
is the effect of stochasticity or noise, which inevitably exists
in a cell and may play important roles \cite{Rao2002}. In this
paper, we advance a probabilistic Boolean network
\cite{Brun2005,Qu2002} on the protein interaction network of yeast
cell cycle. We found that both the biological stationary state and
the biological pathway are well preserved under a wide range of
noise level. When the noise is larger than a value of the order of
the interaction strength, the network dynamics quickly becomes
noise dominating and loses its biological meaning.

\begin{figure}
\centerline{\includegraphics[width=5.5cm]{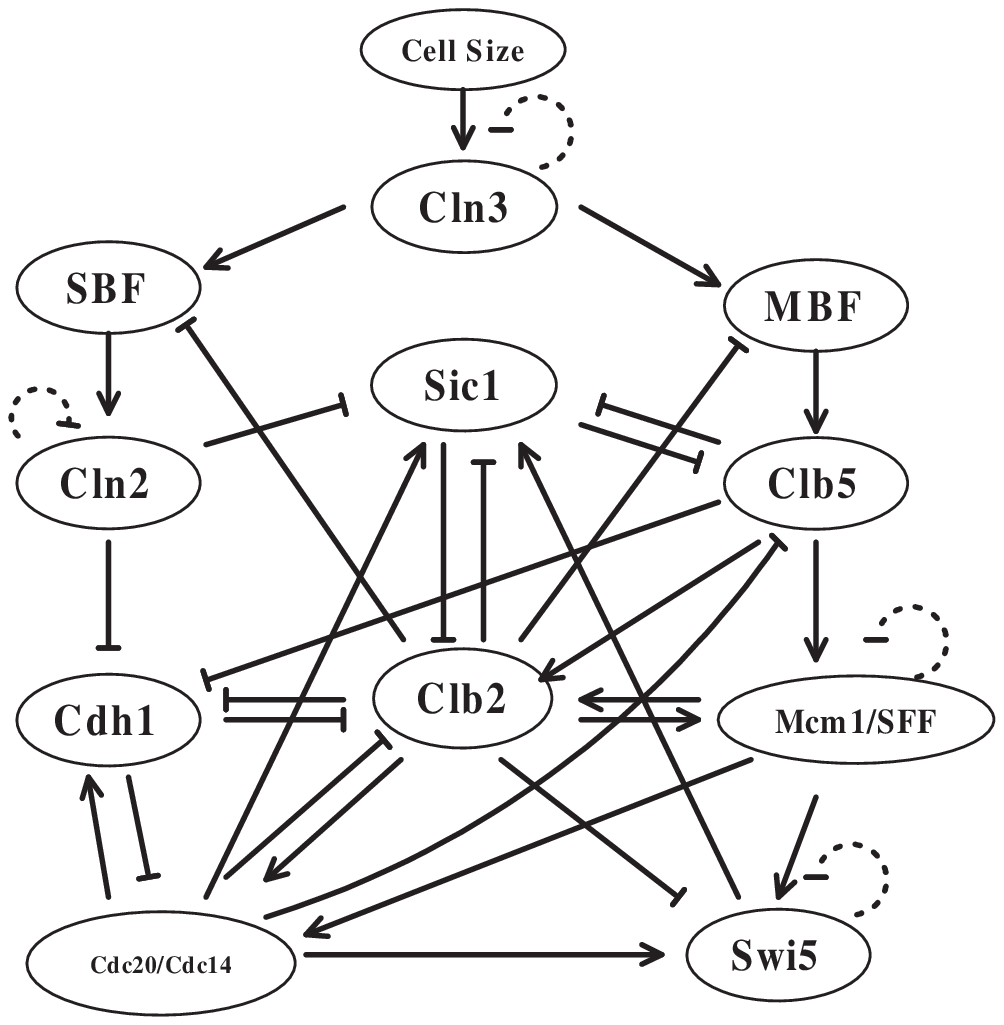}}

 \caption {The cell-cycle network of the budding yeast. Each node represents a
protein or a protein complex. Arrows are positive regulation,
``T"-lines are negative regulation, dotted ``T"-loops are
degradation.} \label{network}
\end{figure}

\section{Method}
\label{Method}

Our stochastic model is based on the updated protein interaction
network of Li \textit{et al.} \cite{Li2004}, which is depicted in
Fig.~\ref{network}. Nodes in the figure represent proteins or
protein complexes. Arrows represent positive interaction, or
``activation". Lines with a bar at the end represent negative
interaction, or ``repression". Dotted loops with a bar represent
self-degradation. We refer the reader to Ref.~\cite{Li2004} for a
full biological account of this network. Here we only give a very
brief summary. There are four phases in the cell cycle process:
the G1 phase in which the cell grows, the S phase in which the DNA
is copied, the G2 phase in which the cell prepares for mitosis,
and the M phase in which the two chromosome copies are separated
and the cell divides into two. There are several checkpoints
during the process to ensure that the next event will not happen
until the current event is finished. So the process could be
blocked at checkpoints. Following Ref.~\cite{Li2004}, we keep only
one such checkpoint in the model: ``cell size". Thus the picture
for the cell division process is the following: The cell is
resting on a stationary state G1 (blocked at the checkpoint until
it grows big enough). The ``signal" to start the cell-cycle
process comes from the ``cell size" which turns on a cyclin Cln3.
Cln3 activates a pair of nodes, SBF and MBF. SBF and MBF stimulate
the transcription of G1/S genes, including those of Cln2 and Clb5.
The S phase cyclin Clb5 initiates DNA replication, after which the
transcription factor complex Mcm1/SFF is turned on, which
stimulates the transcription of many G2/M genes including the gene
of the mitotic cyclin Clb2. The cell will exit from mitosis and
divide into two after Clb2 is being inhibited and degraded by
Cdc20, Cdh1 and Sic1. The cell (or two cells: the mother and the
daughter) now comes back to the stationary G1 state, waiting for
the signal for another round of division. So from a dynamics point
of view, the cell's stationary state G1 is a fixed point. A
``start" signal will take it out of the fixed point, and it will
then go through a specific dynamic trajectory (the biological
pathway for cell division), and come back to the fixed point.

\begin{figure}
\centerline{\includegraphics[width=8.5cm]{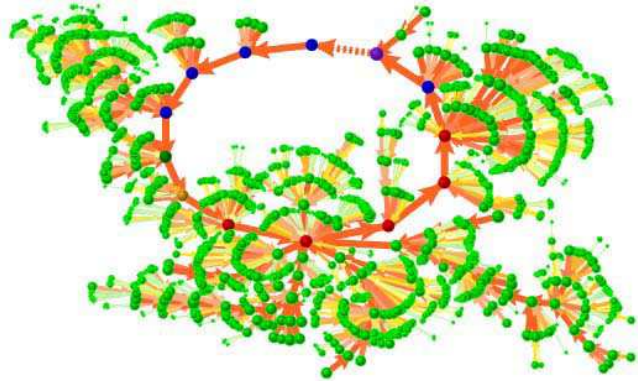}}
\caption {The probability flux. For a node, only the largest flux
from it is shown. The nodes on the biological pathway are denoted
with different colors: purple, the stationary G1 state; blue, the
other G1 states; olive-green, S state; dandelion, G2 state; and red,
M states. All other states are noted in normal green. The
simulations were done with $\alpha =5$ and $\beta =6$.}
\label{biopath1}
\end{figure}

In our model, the 11 nodes in the network shown in
Fig.~\ref{network}, namely, Cln3, MBF, SBF, Cln2, Cdh1, Swi5,
Cdc20, Clb5, Sic1, Clb2, and Mcm1 are represented by variables
$(s_{1},s_{2},...,s_{11})$, respectively. Each node $i$ has only
two values: $s_{i}=1$ and $s_{i}=0$, representing the active state
and the inactive state of the protein $i$, respectively.
Mathematically we consider the network evolving on the
configuration space $S = {\{0, 1\}}^{11}$; the $2^{11}=2048$
``cell states" are labelled by ${\{n=0,1,...,2047\}}$. The
statistical behavior of the cell state at the next time step is
determined by the cell state at the present time step. That is,
the evolution of the network has the Markov property
\cite{Chung1967}. The time steps here are logic steps that
represent causality rather than actual times. The stochastic
process is assumed to be time homogeneous. Under these assumptions
and considerations, we define the transition probability of the
Markov chain as follows:
\begin{eqnarray}
P_r(s_{1}(t+1),...,s_{11}(t+1)|s_{1}(t),...,s_{11}(t))\nonumber\\
=\prod_{i=1}^{11}P_r(s_{i}(t+1)|s_{1}(t),...,s_{11}(t)),
\end{eqnarray}
where
\begin{eqnarray*}
\lefteqn{P_r(s_{i}(t+1)= \sigma_{i}|s_{1}(t),...,s_{11}(t)) = }\\
& & \frac{\exp(\beta(2\sigma_{i}-1)T)}{\exp(\beta T)+\exp(-\beta
T)},
\end{eqnarray*}
if \,
 $T=\sum_{j=1}^{11}a_{ij}s_{j}(t)\ne 0,  \sigma_{i}\in
 \{0,1\}$;
and
\begin{eqnarray}
P_r(s_{i}(t+1)= s_{i}(t)|s_{1}(t),...,s_{11}(t))
=\frac{1}{1+e^{-\alpha}},
\end{eqnarray}
if \,
$T=\sum_{j=1}^{11}a_{ij}s_{j}(t)= 0$.
We define $a_{ij}= 1$ for a positive regulation of $j$ to $i$ and
$a_{ij}= -1$ for a negative regulation of $j$ to $i$. If the
protein $i$ has a self-degradation loop, $a_{ii} = -0.1$. The
positive number $\beta$ is a temperature-like parameter
characterizing the noise in the system \cite{Albeverio1995}.
Noticeably, the actual noises within a cell might not be constant
everywhere, but here we use a system-wide noise measure for
simplicity. To characterize the stochasticity when the input to a
node is zero, we have to introduce another parameter $\alpha$.
This parameter controls the likelihood for a protein to maintain
its state when there is no input to it. Notice that when $\beta$,
$\alpha \rightarrow \infty$, this model recovers the deterministic
model of Li \textit{et al.} \cite{Li2004}. In this case, they
showed that the G1 state (the purple node in Fig.~\ref{biopath1})
is a big attractor, and the path (blue nodes $\rightarrow$
olive-green nodes $\rightarrow$ dandelion nodes $\rightarrow$ red
nodes $\rightarrow$ purple node in Fig.~\ref{biopath1}) is a
globally attracting trajectory. Our study focuses on the
stochastic properties of the system.

Because the Markov chain consists of finite states and is
irreducible, every state is accessible to all others. Therefore
all of the 2048 states constitute a communicated recurrent class,
and the Markov chain is ergodic. In this case, there exists a
probability distribution $\Pi=(\pi_{0},\pi_{1},...,\pi_{2047})$,
an invariant measure, such that for all states $m, n \in
\{0,1,...,2047\}$,
$$\lim_{r \to \infty} p_{mn}(r)=\pi_{n}$$
where $p_{mn}(r)$ is $r$-step transition probability from the
initial state $m$ to the target state $n$.  That is to say, when
$r$ is big enough, the probability for the system to reach state
$n$ is almost independent of the starting position $m$. Even
though each state has a positive probability, the order of the
magnitudes of the probabilities are very different among the
states; some are so small that in realistic cases they can never
be observed.

Our interests are in the asymptotic behavior of the dynamic
system. Steady state probability distribution
$\Pi=(\pi_{0},\pi_{1},...,\pi_{2047})$ can be found by solving
linear equations $\Pi P=\Pi$, where $P$ is the transition matrix
of the Markov chain. The net probability flux from $m$ to $n$ is
then $\pi_{m}p_{mn}-\pi_{n}p_{nm}$, where $p_{mn}$ is the
transition probability from $m$ to $n$. For a given $\beta$, one
can define a pseudo energy for the state $n$ as \cite{Qian1996}
\begin{equation}\label{Dn-energy}
    E_n=-\frac{\log\pi_{n}}{\beta}.
\end{equation}

\begin{figure}
\centerline{\includegraphics[width=8.cm]{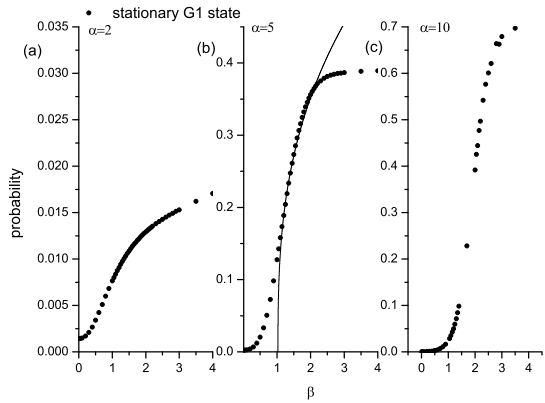}} \caption {The
probability of the stationary G1 state and the biological pathway.
The order parameter $\pi_{G1}$ as a function of $\beta$ with (a),
$\alpha = 2$; (b), $\alpha = 5$, and (c),$\alpha = 10$. The solid
line in (b) is the fitting function
$\varphi(\beta)=0.36|\frac{\beta}{1.03}-1|^{0.36}$.} \label{ratio}
\end{figure}

We first study the property of the biological stationary state G1
and define an ``order parameter" as the probability for the system
to be in the G1 state, $\pi_{G1}$. Plotted in Fig.~\ref{ratio} is
the value of the order parameter ($\pi_{G1}$) as a function of the
control parameter $\beta$ with different $\alpha$. At large
$\beta$ (low ``temperature" or small noise level), the G1 state is
the most probable state of the system and $\pi_{G1}$ has a
significant value. Note that for a finite $\alpha$ there are
always ``leaks" from the G1 state, so that the concept of
attractor in the deterministic model in Ref. \cite{Li2004} cannot
applied here. $\pi_{G1}$ decreases with the decrease of $\beta$;
one observes a transition-like behavior as the function of $\beta$
(similar behavior has been seen in \cite{Qu2002}). In order to
compare this transitional behavior to the transition in the system
of thermodynamic equilibrium, we define
\begin{equation}
\varphi(\beta) = b|\frac{\beta}{\beta_c}-1|^a,
\end{equation}
to fit the order parameter ($\pi_{G1}$) curves in
Fig.~\ref{ratio}, where $\beta_c$, $b$, $a$ are parameters. When
$\alpha$ is fixed to 5, we obtain $\beta_c \approx 1.03$, $b
\approx 0.36$, $a \approx 0.36$ (See Fig.~\ref{ratio}(b)).

At around $\beta_c = 1.03$, $\pi_{G1}$ drops to a very small
value, indicating a ``high temperature" phase in which the network
dynamics cannot converge to the biological steady state G1. The
system is, however, rather resistant to noise. The ``transition
temperature" is quite high: the value of $\beta _c \approx 1.03$
implies that the system will not be significantly affected by
noise until approx $10\%$ of the updating rules are wrong
($e^{-1.03}/(e^{1.03}+e^{-1.03}) \approx 0.1$).

We next study the statistical properties of the biological pathway
of the cell-cycle network. We search the probability for the
system to be in any of the biological states along the biological
pathway, as a function of $\beta $. One observes a similar
transition-like behavior as shown in Fig.~\ref{ratio}. The jump of
the probability of the states along the biological pathway in the
low temperature phase is due to the fact that in this phase the
probability flux among different states in the system is
predominantly the flow along the biological pathway. To visualize
this, we show in Fig.~\ref{biopath1} an example of the probability
flux among all 2048 states. Each node in Fig.~\ref{biopath1}
represents one of the 2048 states. The size of a node reflects the
stationary distribution probability of the state. If the
stationary probability of a state is larger than a given threshold
value, the size of the node is in proportion to the logarithm of
the probability. Otherwise, the node is plotted with the same
smallest size. The arrows reflect the net probability flux (only
the largest flux from any node is shown). The probability flux is
divided into seven grades, which are expressed by seven colors:
light-green, canary, goldenrod, dandelion, apricot, peach and
orange. The warmer the color is,  and the wider the arrow is, the
larger the probability flux. The width of an arrow is in
proportion to the logarithm of the probability flux it carries.
The arrow representing the probability flux from the stationary G1
state to the excited G1 state (the START of the cell-cycle) is
shown in dashed lines. One observes that once the system is
``excited" to the START of the cell cycle process (here by noise
($\alpha$) and in reality mainly by signals like ``cell size") the
system will essentially go through the biological pathway and come
back to the G1 state. Another feature of Fig.~\ref{biopath1} is
that the probability flux from any states other than those on the
biological pathway is convergent onto the biological pathway.
Notice that this diagram also characterizes the properties of
fixed points that are ignored by Li (Ref.~\cite{Li2004}). Those
fixed points also converge onto biological pathway. For $\beta <
\beta_c$, this feature of a convergent high flux bio-pathway
disappears.

\begin{figure}
\centerline{\includegraphics[width=8.5cm]{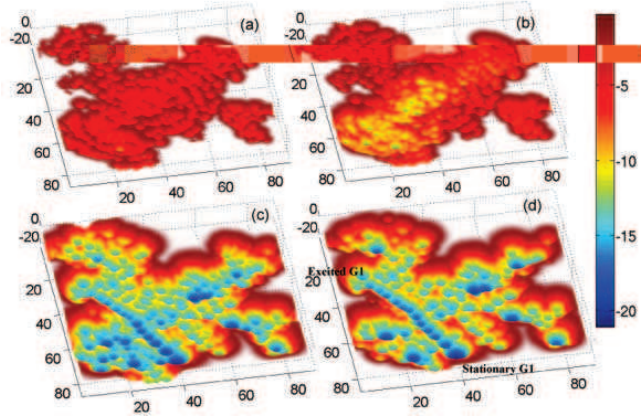}} \caption {The
``potential" landscape of the system before and after the critical
point. (a) $\beta = 0.01$, (b) $\beta = 0.6$, (c) $\beta =1.5$,
(d) $\beta =6.0$, all for $\alpha =5$. The color code gives the
relative value of the potential function.} \label{3Dentropy}
\end{figure}

In the previous discussions, we see that there is a ``phase
transition" as a function of the ``temperature" in the stochastic
cell-cycle model. The next step is to try to understand this
transition-like behavior. For this purpose, we define a ``potential"
function and study the change of the ``potential landscape" as a
function of $\beta$. Specifically, we define
\begin{equation}\label{entropy}
    S_n=-\log\pi_{n}=\beta E_n,
\end{equation}
where $E_n$ is the pseudo energy defined in Eq.~[\ref{Dn-energy}].
Fig.~\ref{3Dentropy} shows four examples of $\Delta S_n = S_n -
S0$ distribution, where the reference potential $S0$ in each plot
is set as the highest potential point in the system. Note that the
11-dimensional phase space is reduced to two dimensions and there
is no distance metric among the states in the 2D phase space. The
states in 2D are arranged for clarity, which reflect a kind of
dynamic relationship as in Fig.~\ref{biopath1}. One observes that
far from the critical point ($\beta = 0.01$,
Fig.~\ref{3Dentropy}(a)), the potential values are high (around
-4), and the landscape is flat. Near but below the critical point
($\beta = 0.6$, Fig.~\ref{3Dentropy}(b)), some local minima (blue
points) become more pronounced, but the landscape still remains
rather flat. We notice that these minimum points do not correspond
to the biological pathway. Right after the critical point ($\beta
=1.5$, Fig.~\ref{3Dentropy}(c)), the system quickly condenses into
a landscape with deep valleys \cite{Waddington1957}. The state
with the lowermost potential value corresponds to the stationary
G1 state. A linear line of blue dots from up-left to down-middle
corresponds to the biological pathway, which forms a deep valley.
Some deep blue dots out of the biological pathway are local
attractors in Ref.~\cite{Li2004}. Notice that although their
potential values are low, they attract only a few nearby initial
states--all these points are more or less isolated. After the
critical point, the potential landscape does not change
qualitatively (see Fig.~\ref{3Dentropy}(d) with $\beta=6$). As
$\beta$, $\alpha \rightarrow \infty$, the landscape becomes nine
deep holes, each corresponding to an attractor of the determinate
system.

\section{Conclusion}
\label{Conclusion}

In conclusion, we introduced a stochastic model for the yeast cell
cycle network. We found that there exists a transition-like behavior
as the noise level is varied. With large noise, the network behaves
randomly; it cannot carry out the ordered biological function. When
the noise level drops below a critical value, which is of the same
order as the interaction strength ($\beta_c \approx 1.03$), the
system becomes ordered: the biological pathway of the cell cycle
process becomes the most probable pathway of the system and the
probability of deviating from this pathway is very small. So in
addition to the dynamical and the structural stability
\cite{Li2004}, this network is also stable against stochastic
fluctuations. We used a pseudo potential function to describe the
dynamic landscape of the system. In this language, the biological
pathway can be viewed as a valley in the landscape
\cite{Ao2004,Zhu2004,Waddington1957}. This analogy to equilibrium
systems may not be generalizable, but it would be interesting to see
if one can find more examples in other biological networks, which
are very special dynamical systems.

\section{Acknowledgments}
\label{Acknowledgments}

We thanks W.Z. Ma for preparing the Fig.~\ref{3Dentropy} and H. Yu
for helpful discussion. This research is supported by the grants
from National Natural Science Foundation of China (No. 90208022,
No. 10271008, No. 10329102), the National High Technology Research
and Development of China (No. 2002AA234011) and the National Key
Basic Research Project of China (No. 2003CB715903)






\bibliographystyle{elsart-num}
\bibliography{reference_phyd_revised}

\end{document}